\documentclass{aa} 
\usepackage{graphics} 
\usepackage{amsmath}
\input{psfig} 

\begin{document}

\title{The balance between shocks and AGN photoionization  in radio sources and its relation to the radio size}

\subtitle{}

\author{E. Moy \inst{1,2} \and B. Rocca-Volmerange \inst{1,3}}

\offprints{moy@fast.mpe-garching.mpg.de}

\institute{
\inst{1} Institut    d'Astrophysique de  Paris,    98\,{\it  bis}, Boulevard Arago, F-75014 Paris, France. \\
\inst{2} Max-Planck-Institut f\"ur extraterrestrische Physik, Postfach 1312, D-85741 Garching, Germany. \\
\inst{3} Institut d'Astrophysique Spatiale, B\^at. 121, Universit\'e Paris XI, F-91405 Orsay, France.\\
}
\authorrunning{Moy \& Rocca-Volmerange}

\titlerunning{Shocks and photoionization in AGNs}

\date{}

\abstract{We have analyzed the ultraviolet and optical emission line ratios of a large sample of extragalactic  radio sources (QSOs and radio galaxies), with the help of models combining AGN photoionization and shocks. The results strongly suggest that the two ionizing mechanisms frequently coexist. The model sequences obtained by varying the  balance between shocks and AGN photoionization account for most emission line data in the 12 line ratio diagrams we have considered. In the frequently used diagrams involving [OIII]$\lambda$5007/H$\beta$, [OI]$\lambda$6300/H$\alpha$, [NII]$\lambda$6584/H$\alpha$ and [SII]$\lambda\lambda$6716,6731 (Veilleux \& Osterbrock 1987), the effect of varying the shock-photoionization balance mimics a variation of the ionization parameter ($U$) in traditional photoionization sequences. In most of the remaining diagrams, such as  [OI]$\lambda$6300/[OIII]$\lambda$5007 vs. [OIII]$\lambda$4363/[OIII]$\lambda$5007,   [OIII]$\lambda$5007/H$\beta$ vs. [OIII]$\lambda$4363/[OIII]$\lambda$5007  and   CIII]$\lambda$1909/[CII]$\lambda$2326 vs.  CIV$\lambda$1549/[CII]$\lambda$2326, the data can {\it only} be accounted for if both photoionization and shocks contribute to the line fluxes. The coexistence of shocks and AGN photoionization also provides an explanation for the most extreme objects in the NV$\lambda$1240/HeII$\lambda$1640 vs. NV$\lambda$1240/CIV$\lambda$1549 diagram without requiring largely super-solar metallicities.
In addition, we  show that there is a relationship  between the [OII]$\lambda$3727/[OIII]$\lambda$5007 ratio (i.e., the ionization level of the gas) and the radio size in radio galaxies.  This strongly supports the hypothesis that the most compact ($\le$  2kpc) and the largest ($\ge$ 150 kpc) sources are dominated by photoionization, while intermediate-sized radio galaxies are dominated by shocks.  We briefly discuss the possible origin of the relation between the shock-ionization balance and the radio size.
\keywords{Line: formation -- Galaxies: active -- Galaxies: evolution --
Galaxies: ISM -- Quasars: emission lines
               }
}

\maketitle

%

\section{Introduction}
One  of the  most  debated  questions about  the  interaction between active galactic nuclei (AGN) and their environment is the nature of the ionizing source of the extended structures  of
ionized gas detected in powerful radio galaxies (Spinrad  \& Djorgovski 1984;
Tadhunter et al. 1988; Baum et  al.  1988)  and QSOs (Hes et al. 1996). The emission line spectra of these regions impose the strongest constraints on this issue. Initial studies of extended line emission in  radio sources showed that photoionization models provide an  acceptable fit of emission line ratios in the optical (Robinson et al. 1987; Saunders et al. 1989;  Prieto et al. 1993) and the UV (Villar-Mart\'{\i}n et al. 1997), in good agreement with the  ``unified scheme'' (e.g. Antonucci 1993).

However, some properties of the extended emission cannot be accounted for by the photoionization hypothesis. Some AGN have a spectrum dominated by low ionization lines (e.g.  Koski \& Osterbrock 1976; Durret 1990), or show [OIII]$\lambda$4363/[OIII]$\lambda$5007 ratios indicative of high electron temperatures ($T_\mathrm{e}$ ~$\sim$ 20000 K; Ferland \& Osterbrock 1987). These two properties are more reminiscent of ionizing shocks. Shocks can be produced, for example, by the interactions of the radio ejecta with the interstellar medium (ISM), as  suggested by the spatial coincidence between radio and line emission in radio sources (Baum \& Heckman 1989; McCarthy et al. 1995; Ridgway \& Stockton 1997; Lehnert et al. 1999). The disturbed kinematics of radio galaxies are also indicative of the influence of shocks in the ionizing process (Villar-Mart\'{\i}n et al. 1999a). With the help of theoretical models (Dopita \& Sutherland 1996), evidence of ionizing shocks has been found in an increasing number of AGN from the analysis of their line ratios (Sutherland et al. 1993;  Clark et al. 1998; Koekemoer \& Bicknell 1998; Villar-Mart\'{\i}n et al. 1999c). 

Despite their success, however, shock models remain  on average  less efficient than photoionization models in explaining the emission line properties of  AGN, especially in the UV (Villar-Mart\' \i n  et al.  1997; De Breuck et al. 2000). Therefore, shocks should probably not be regarded as an alternative to AGN photoionization, but more likely as an additional  ionization source that may  be dominant only under some circumstances. In this paper we investigate the possibility that shocks induced by the radio ejecta and AGN photoionization coexist in QSOs and radio galaxies. 

To test this hypothesis, we systematically compare observed emission line ratios, both in the UV and in the optical, with models combining the two ionization processes in varying proportions. The shock models are computed with MAPPINGSIII (Dopita \& Sutherland 1996) and the photoionization models with CLOUDY (Ferland 1996).  Our approach implicitly assumes that shocks and photoionization don't occur simultaneously in individual clouds. This is  a reasonable approximation in the case of jet-induced shocks located near radio hot spots (Clark et al. 1998; Villar-Mart\'{\i}n et al. 1999b). At such distances from the nucleus, the AGN radiation is weak due to geometrical dilution, and the ionization by shocks should be dominant. Conversely, clouds lying closer to the AGN should be mainly photoionized.  However, our models probably don't apply in the case of shocks induced by inflows or outflows (see Contini \& Aldrovandi 1983; Contini \& Aldrovandi 1986; Aldrovandi \& Contini 1984, 1985; Contini \& Viegas-Aldrovandi 1987; Viegas-Aldrovandi \& Contini 1989).

We also discuss in detail the relationship between the radio size and the role of shocks in the ionizing process. The present work confirms the conclusion of Best et al. (2000b) that large radio sources are dominated by AGN photoionization. Here we show that compact sources are {\it also} dominated by  AGN photoionization. Shocks play a major role only in some of the intermediate-sized radio sources.

Sect. 2 presents our modeling of the ionizing processes with the help of  CLOUDY and MAPPINGSIII. The data set used for the comparison with the model predictions is described in Sect. 3. The results of this comparison are presented in Sect. 4, with a particular emphasis on the NV$\lambda$1240/HeII$\lambda$1640 vs.
NV$\lambda$1240/CIV$\lambda$1549  diagram,  because of its importance as a metallicity diagnostic. In Sect. 5, we discuss the evidence for ionizing shocks in radio sources, and investigate the possible origin of  
the link between the shock-photoionization balance and the radio size.  We summarize our  conclusions in Sect. 6.

\section{Modeling the ionizing processes}

\subsection{Photoionization by the central AGN}

AGN photoionization models (hereafter A) are computed with  CLOUDY (Ferland 1996). The emission line spectra of photoionized materials depend on the slope of the ionizing continuum,  on the  ionization parameter $U$\footnote{defined as $U$ = $F^\mathrm{H_{0}}$/$n_\mathrm{H} c$, where $F^\mathrm{H_{0}}$ is the ionizing flux and $n_\mathrm{H}$ the hydrogen density.},  on the gas density and on its metallicity.
  The ionizing continuum follows a power law ($\Phi_{\nu}$ $\propto$ $\nu^{\alpha}$) truncated at 0.01 and 20 Ryd. 

Two values of the spectral index are considered:  $\alpha$ = -1, in agreement with the analysis of UV line ratios in high-$z$ radio galaxies (Villar-Mart\'{\i}n et al. 1997), and $\alpha$ = -1.5, as suggested  by optical line ratios in low-$z$ sources (Robinson et al. 1987). The ionization parameter  varies between log\,$U$ =  -4 and log\,$U$ = -1. We consider two values for the hydrogen density: $n_\mathrm{H}$ = 10$^{2}$  cm$^{-3}$  and $n_\mathrm{H}$ = 10$^{6}$ cm$^{-3}$. The first value is typical of extended emission line regions (McCarthy et al. 1990).  The secund one accounts for a possible contamination of UV lines by denser regions, as suggested by Villar-Mart\'{\i}n et al. (1999b). Two different metallicities, $Z$ = $Z_{\odot}$ and $Z$ = 1.5 $Z_{\odot}$, are considered. The  abundances are as in Moy et al. (2001), except for nitrogen: in the present work the evolution of the N/O ratio  with $Z$ follows the empirical relation proposed by Bresolin et al. (1999). 

\subsection{Ionizing shock}

We use MAPPINGSIII (Dopita \& Sutherland 1996)  to compute the emission line spectra emitted by shocked materials. The precursor density is 1 cm$^{-3}$, as in Dopita \& Sutherland (1995). We considered  shock velocities $v$ between  100  and 1000 km s$^{-1}$. The magnetic parameter B/$\sqrt(n)$ is fixed to 3 in all calculations. This value corresponds to equipartition between thermal and magnetic pressures (Dopita \& Sutherland 1996). For the sake of consistency, metallicities ($Z_{\odot}$ and1.5 $Z_{\odot}$) and chemical abundances are exactly those adopted for photoionization calculations.  We consider both pure shock (hereafter S) and shock+precursor (hereafter SP) models.

The upward spectrum produced by a $v$ = 1000 km s$^{-1}$ shock is too hard to be fully absorbed by the precursor region. To construct an appropriate SP model for $v$ = 1000 km s$^{-1}$, the precursor model is therefore truncated at  $F_\mathrm{HII}$ $\sim$ 0.7, where $F_\mathrm{HII}$ is the fraction of ionized hydrogen, so that it is ``matter-bounded'' (Viegas \& Prieto 1992).

\begin{center}
\begin{figure*}[htbp]
\begin{center}
\centerline{\hbox{
\psfig{figure=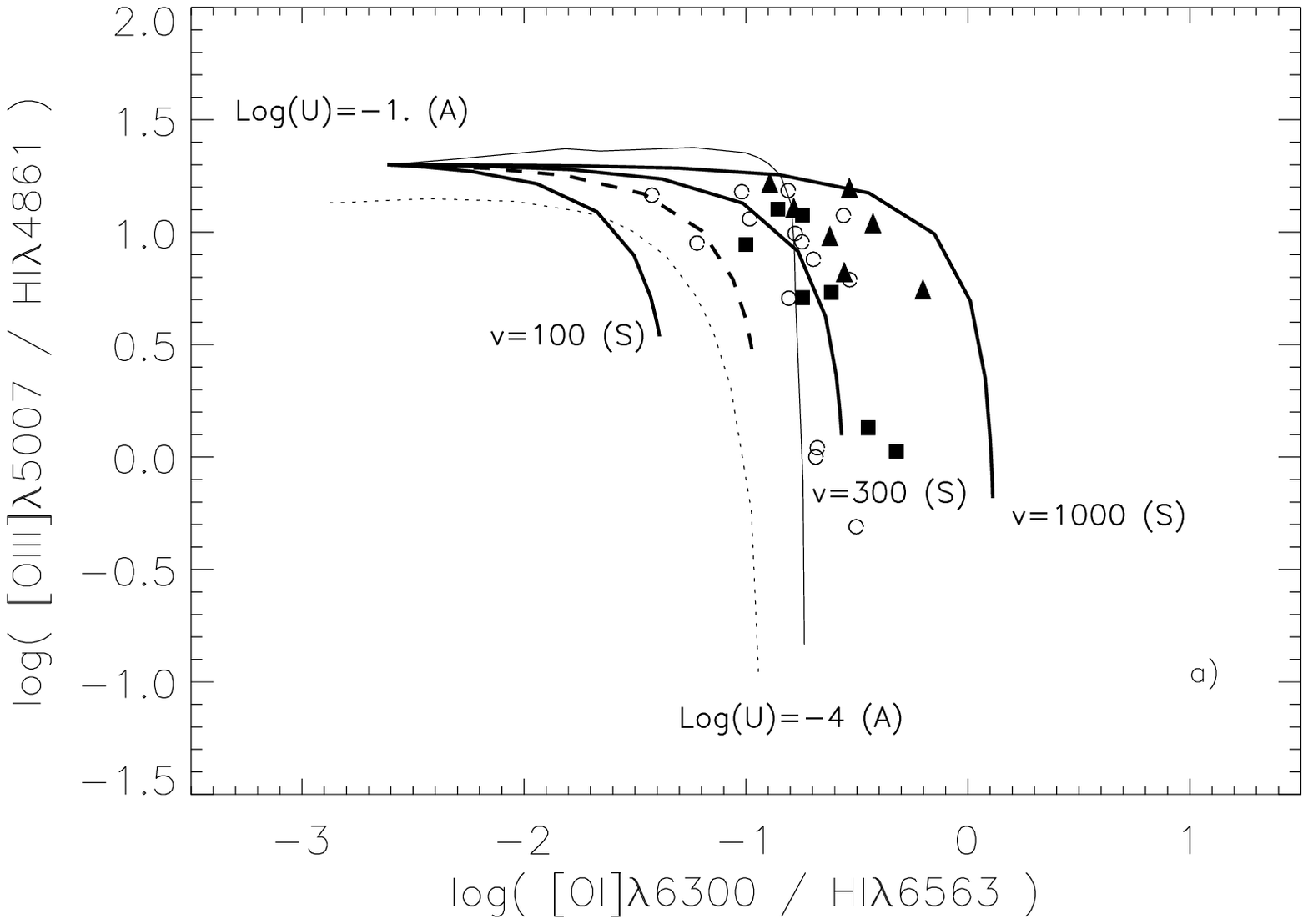,bbllx=30pt,bblly=0pt,bburx=480pt,bbury=
350pt,height=6cm}
\psfig{figure=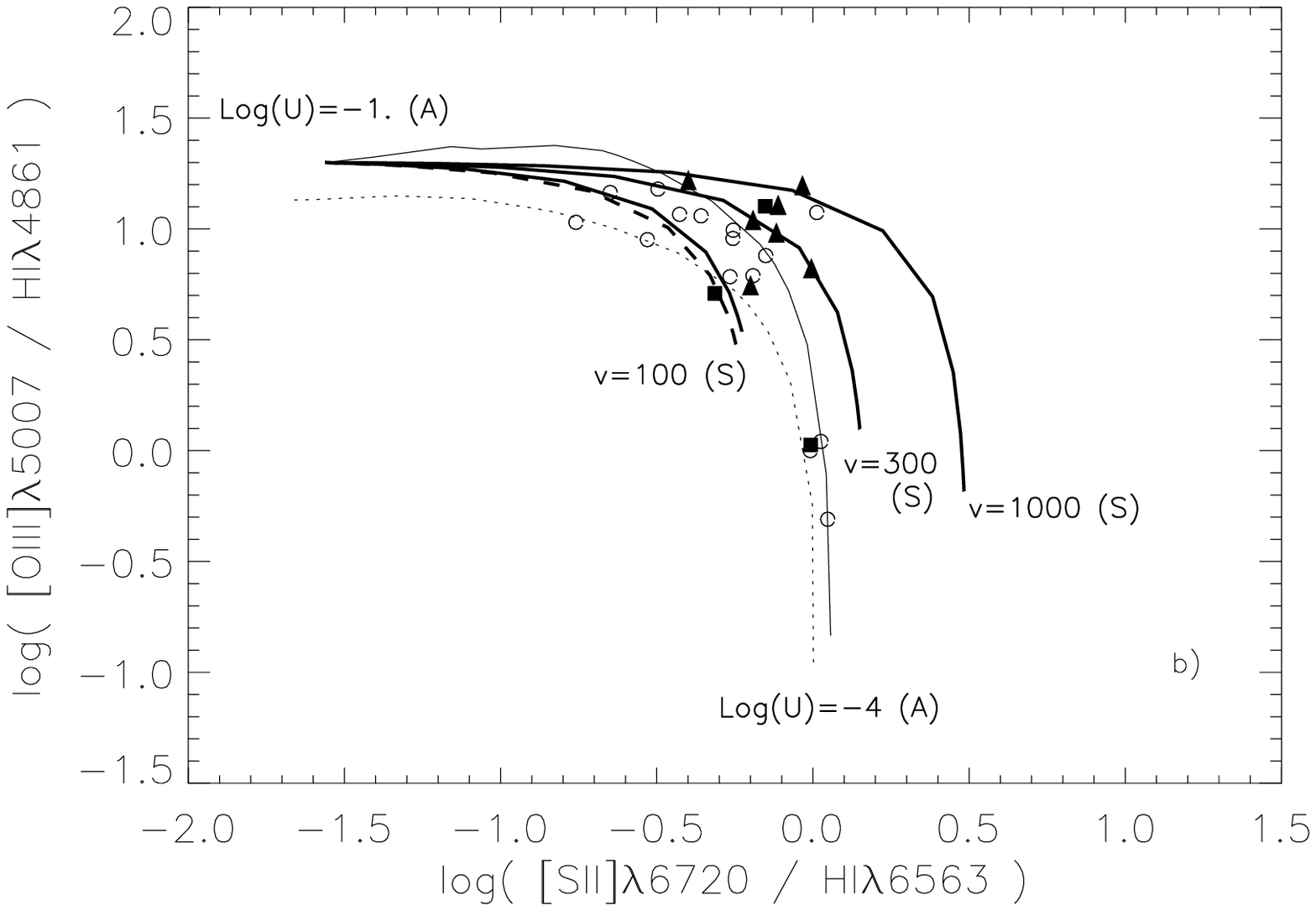,bbllx=0pt,bblly=0pt,bburx=480pt,bbury=
350pt,height=6cm}
}}
\centerline{\hbox{
\psfig{figure=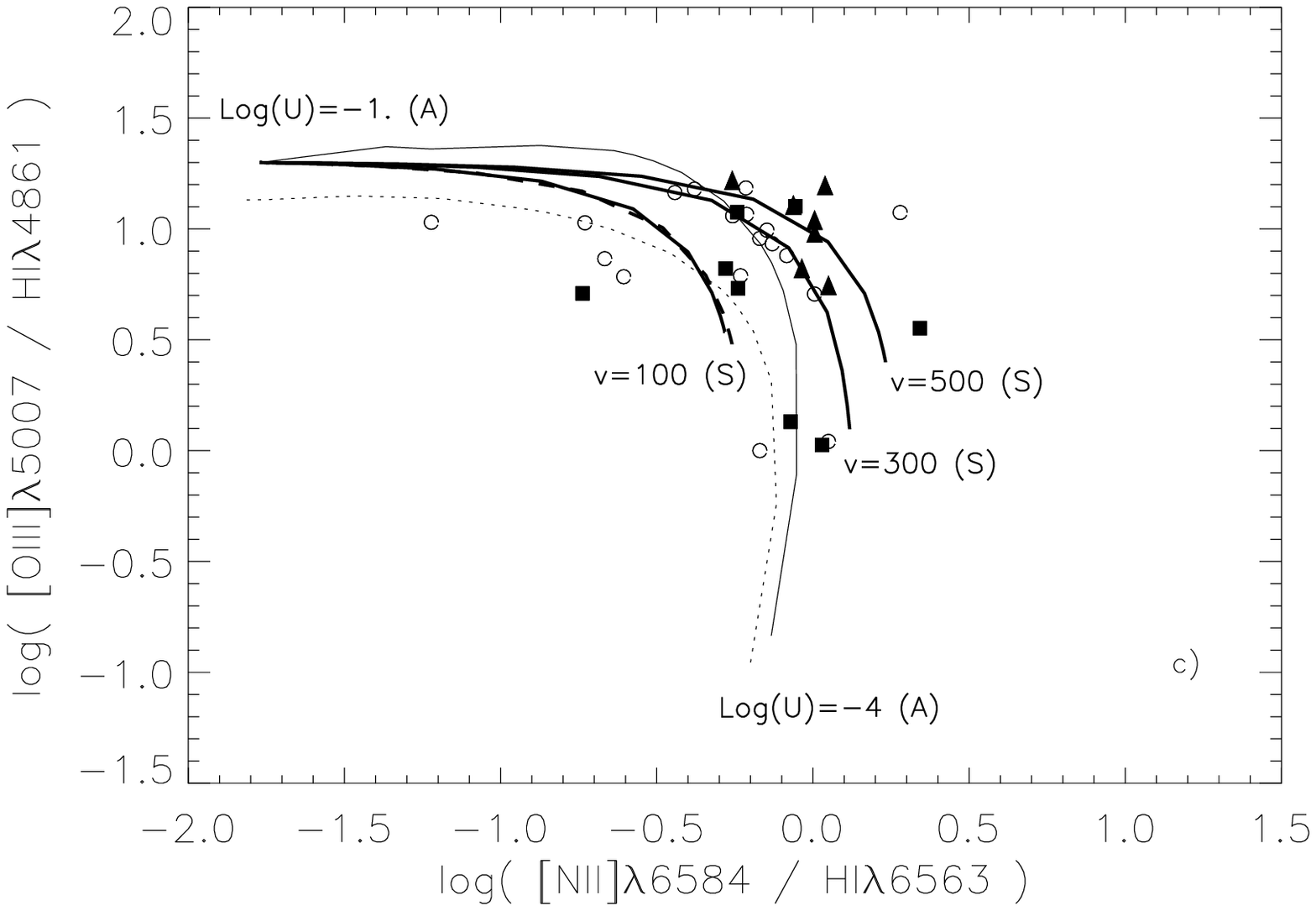,bbllx=30pt,bblly=0pt,bburx=480pt,bbury=
350pt,height=6cm}}}
\parbox{17.cm}{\caption{Comparisond between the predictions of two component A+S models (thick solid lines) and observational data in the three diagrams of Veilleux \& Osterbrock (1987). The metallicity is solar, and the density $n_\mathrm{H}$ is equal to 100 cm$^{-3}$. The values of log\,$U$ and $v$ are indicated on the plot. Two photoionization model sequences (-4 $\le$ log\,$U$ $\le$ -1),  with spectral index $\alpha$ = -1 (thin solid line) and $\alpha$ = -1.5 (dotted line), are  shown for comparison. An example of an A+SP model sequence with log\,$U$ = -1 and $v$ = 300 km s$^{-1}$ is also plotted (dashed line). Filled squares indicate emission line ratios for spatially integrated (nuclear+extended) emission line regions, filled triangles for nuclear regions only, and open circles for extended regions only. Note the similarity between the A+S sequences and the pure photoionization sequences.}}
\end{center}
\end{figure*}
\end{center}

\subsection{Combining photoionization and shock models}

A two-component model is obtained by combining the results of a shock model (characterized by the shock velocity $v$) and of a photoionization model (characterized by the ionization parameter $U$). Only models with similar metallicities are combined in this way. The total flux $F_{i}$ in a given line $i$ is computed as follows:

\begin{equation}
F_{i} = A_\mathrm{A/S} \times  F_{i}^{v} + (1-A_\mathrm{A/S}) \times F_{i}^{U} \ \ A_\mathrm{A/S} \in [0,1]
\end{equation}

\noindent where $F_{i}^{v}$ and $F_{i}^{U}$ are the fluxes predicted by the shock and  photoionization models, respectively. By varying the $A_\mathrm{A/S}$ parameter between 0 and 1, we obtain full sequences between the ``pure shock'' and the ``pure photoionization'' cases.

\section{Data sources}

Our data compilation, listed in the Appendix (Table A.1), includes emission lines for  369 radio galaxies. 122  have a Fanaroff \& Riley (1974) classification (10 FR1s and 112 FR2s). The sample also  includes 23 compact steep spectrum (CSS) objects  and 59 QSOs.  For objects with both narrow and broad lines, only the narrow component is considered.  We classify the observations according to their spatial coverage: nuclear (including only the central regions), extended (including only the off-nuclear regions), and spatially integrated (covering both the nuclear and extended regions). We classified the observations of Simpson et al. (1996) as ``spatially integrated''  in spite of their limited spatial coverage, since they include at least part of the extended emission.  The radio sizes  are estimated from the apparent largest angular sizes (LAS), assuming $\Omega_{0}$ = 1, $\Lambda_{0}$ = 0 and $H_{0}$ = 50 km s$^{-1}$ Mpc$^{-1}$. The source papers for the LAS are listed in column 3 of Table A1. For sources from Morganti et al. (1993), the LAS were directly measured from the radio images.

\begin{center}
\begin{figure*}[htbp]
\begin{center}
\centerline{\hbox{
\psfig{figure=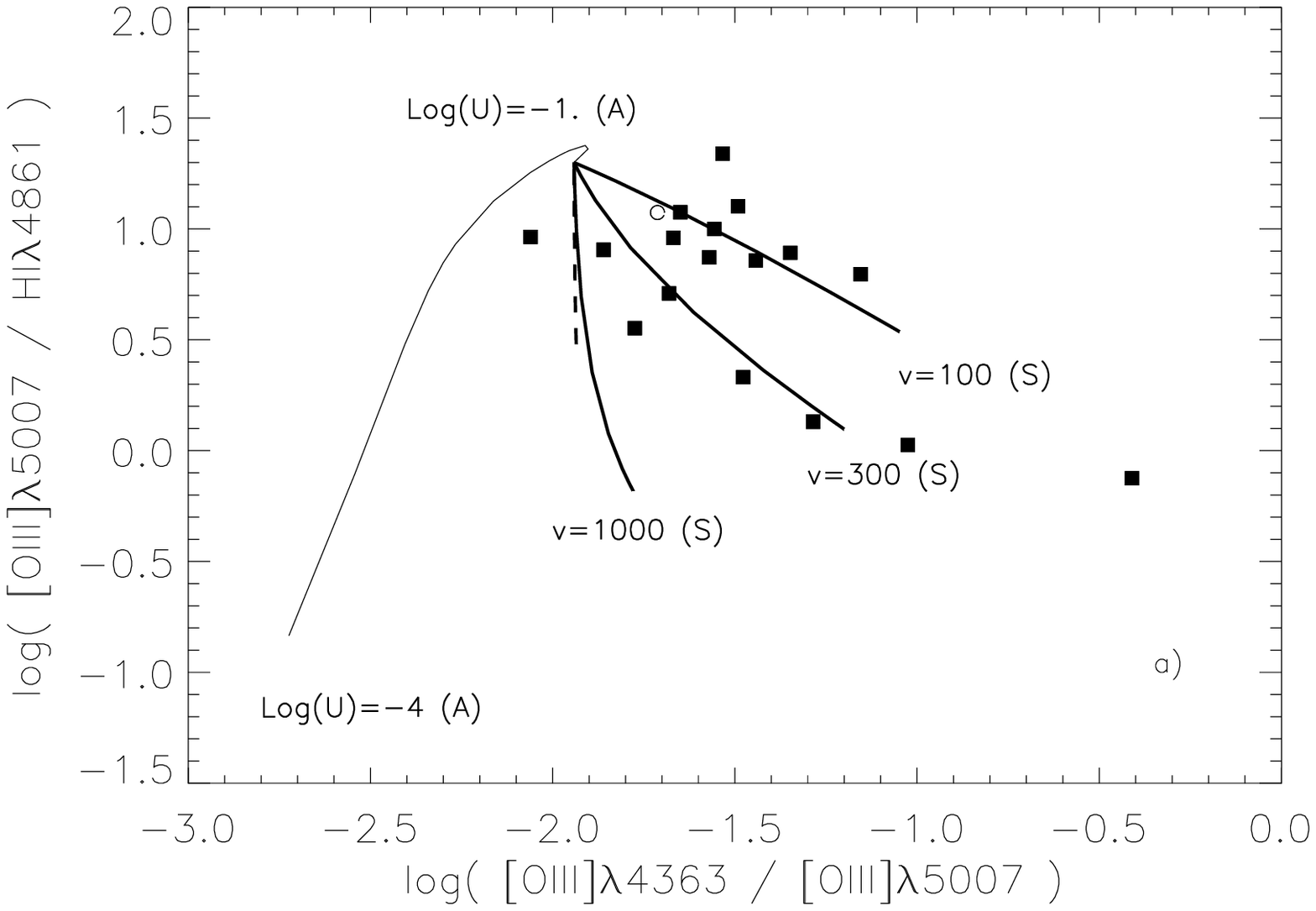,bbllx=30pt,bblly=0pt,bburx=480pt,bbury=
350pt,height=6cm}
\psfig{figure=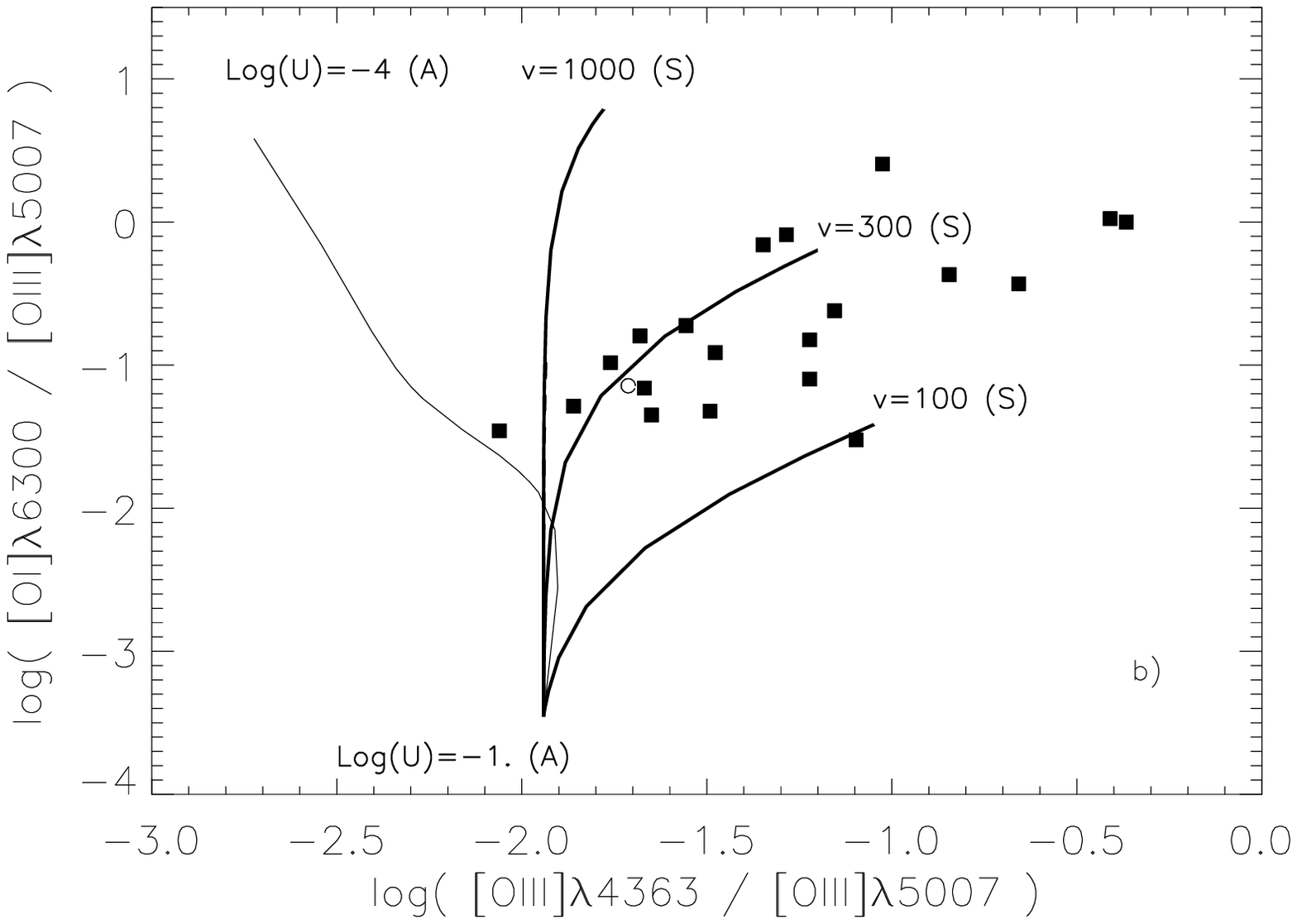,bbllx=0pt,bblly=0pt,bburx=480pt,bbury=
350pt,height=6cm}}}
\parbox{14.cm}{\caption{[OIII]$\lambda$5007/H$\beta$  and  [OI]$\lambda$6300/[OIII]$\lambda$5007 
vs. [OIII]$\lambda$4363/[OIII]$\lambda$5007. Symbols as in Fig. 1.  The photoionization sequence with $\alpha$ = -1.5 is not plotted. Remaining curves as in Fig. 1: thick solid lines are A+S models; thin solid line are pure  photoionization models  (-4 $\le$ log\,$U$ $\le$ -1),  with spectral index $\alpha$ = -1; dashed line are A+SP models. The data trends and dispersions are compatible with A+S models. The A+SP sequence (dashed line) agrees almost exactly  with the $v$ = 1000 km s$^{-1}$ A+S sequence. Pure photoionization models are clearly excluded.}}
\end{center}
\end{figure*}
\end{center}

\section{Diagnostic diagrams with combined models}

Figures 1 to 6  present the comparisons between the model predictions and our compilation of line ratios.  Since large ranges of ionization parameter and shock velocity are
considered, the number of $A_\mathrm{A/S}$ sequences is very high.  For the sake of clarity, only a few sequences, characterized by the parameter pair (log\,$U$, $v$),  are plotted on each figure.  These are (log\,$U$,$v$) = (-1,300), (-1,100), and (-1,1000) for $A_\mathrm{A/S}$ sequences combining one photoionization model with one pure shock model, hereafter A+S sequences (thick solid lines). The first of these sequences provides a good fit to the data in many optical diagrams and is therefore regarded as the ``reference'' sequence. The latter two sequences  usually define the borders of the area covered by  two-component models. 

One example sequence which includes the emission from a precursor (hereafter A+SP sequence), with (log\,$U$,$v$) = (-1,300), is also plotted (dashed line). A ``classical'' AGN photoionization model sequence with spectral index $\alpha$ = -1 and -4 $\le$ log\,$U$ $\le$ -1 is shown for comparison (thin line). If not specified, the metallicity is always solar, and the hydrogen density  $n_\mathrm{H}$ is always  100 cm$^{-3}$. Some diagrams also plot an additional model sequence (dotted line) with different parameters, whose values depend on the line ratio under consideration (see below).

\subsection{Optical line ratios}

\subsubsection{Veilleux and Osterbrock (1987) diagrams}

 Figure 1 shows the comparisons between the various models and our compilation of data in the three diagnostic diagrams of Veilleux \& Osterbrock (1987): [OIII]$\lambda$5007/H$\beta$ vs. [OI]$\lambda$6300/H$\alpha$ (Fig. 1a), [OIII]$\lambda$5007/H$\beta$ vs. [SII]$\lambda$6720/H$\alpha$ (Fig. 1b) and [OIII]$\lambda$5007/H$\beta$ vs. [NII]$\lambda$6584/H$\alpha$ (Fig. 1c). On each graph,  the A+S sequences (thick solid lines) are very similar to the  pure AGN photoionization sequence (thin line). This result demonstrates that  a variation of the  balance between shocks and AGN photoionization mimics a variation of the ionization parameter $U$, which has long been the traditional way of interpreting optical line ratio diagrams.
\begin{center}
\begin{figure}[thbp]
\begin{center}
\centerline{\hbox{
\psfig{figure=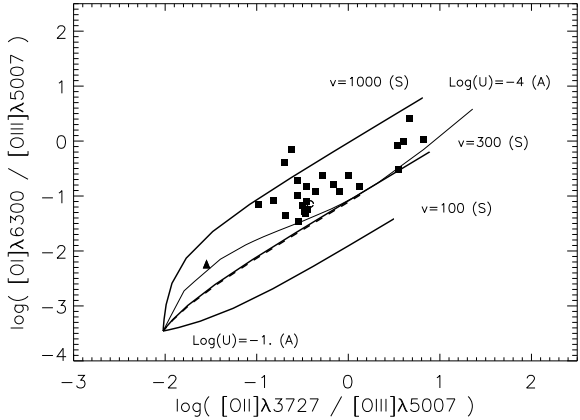,bbllx=0pt,bblly=0pt,bburx=550pt,bbury=350pt,height=6cm}}}
\parbox{8.cm}{\caption{[OI]$\lambda$6300 / [OIII]$\lambda$5007      
vs. [OII]$\lambda$3727/ [OIII]$\lambda$5007. Symbols  as in Fig. 1, and curves as in
Fig. 2.  The A+SP (dashed line)  sequence is superimposed to  the $v$ =
300 km s$^{-1}$  A+S sequence. As in Fig. 1,  the A+S sequences (thick
lines)  are very similar  to the  pure photoionization  sequence (thin
line),       but       with        a       larger       range       of
[OI]$\lambda$6300/[OIII]$\lambda$5007.}}
\end{center}
\end{figure}
\end{center}
The $\alpha$ = -1 photoionization models (thin solid line) underpredict the observed [OI]$\lambda$6300/H$\alpha$, [SII]$\lambda$6720/H$\alpha$ and [NII]$\lambda$6584/H$\alpha$ ratios, sometimes by as much as 0.5 dex. In the case of  $\alpha$ = -1.5 (dotted line),  the situation is even worse, due to the lack of high-energy photons and the subsequent reduction of the ``partially ionized zone''. In order to reconcile pure photoionization models with the data, one could in principle  assume either a lower metallicity (see Ferland \& Netzer 1983) or a higher density. However, the first hypothesis does not solve the discrepancy for [NII]$\lambda$6584/H$\alpha$, as already emphasized by Ferland \& Netzer (1983) and  Veilleux \& Osterbrock (1987). The second hypothesis is ruled out by the high [SII]$\lambda$6720/H$\alpha$ ratios observed in our sample. Due to collisional de-excitation, the sulphur lines would be weaker than observed if densities were higher  than  $n_\mathrm{H}$ = 100 cm$^{-3}$.  A+S models (combining  AGN photoionization and pure shocks;  thick lines)  account for most of the  data. A+SP models (combining AGN photoionization and shocks+precursors; dashed line) are acceptable as well. The sequence with (log\,$U$,$v$) = (-1,300) provides the best fit to the data on the three diagrams.  The dispersion around this sequence, especially in [OI]$\lambda$6300/H$\alpha$ (Fig. 1a) and [SII]$\lambda$6720/H$\alpha$ (Fig. 1b), requires that the shock velocity varies between 100 and 1000 km s$^{-1}$. 

\begin{center}
\begin{figure*}[htbp]
\begin{center}
\centerline{\hbox{
\psfig{figure=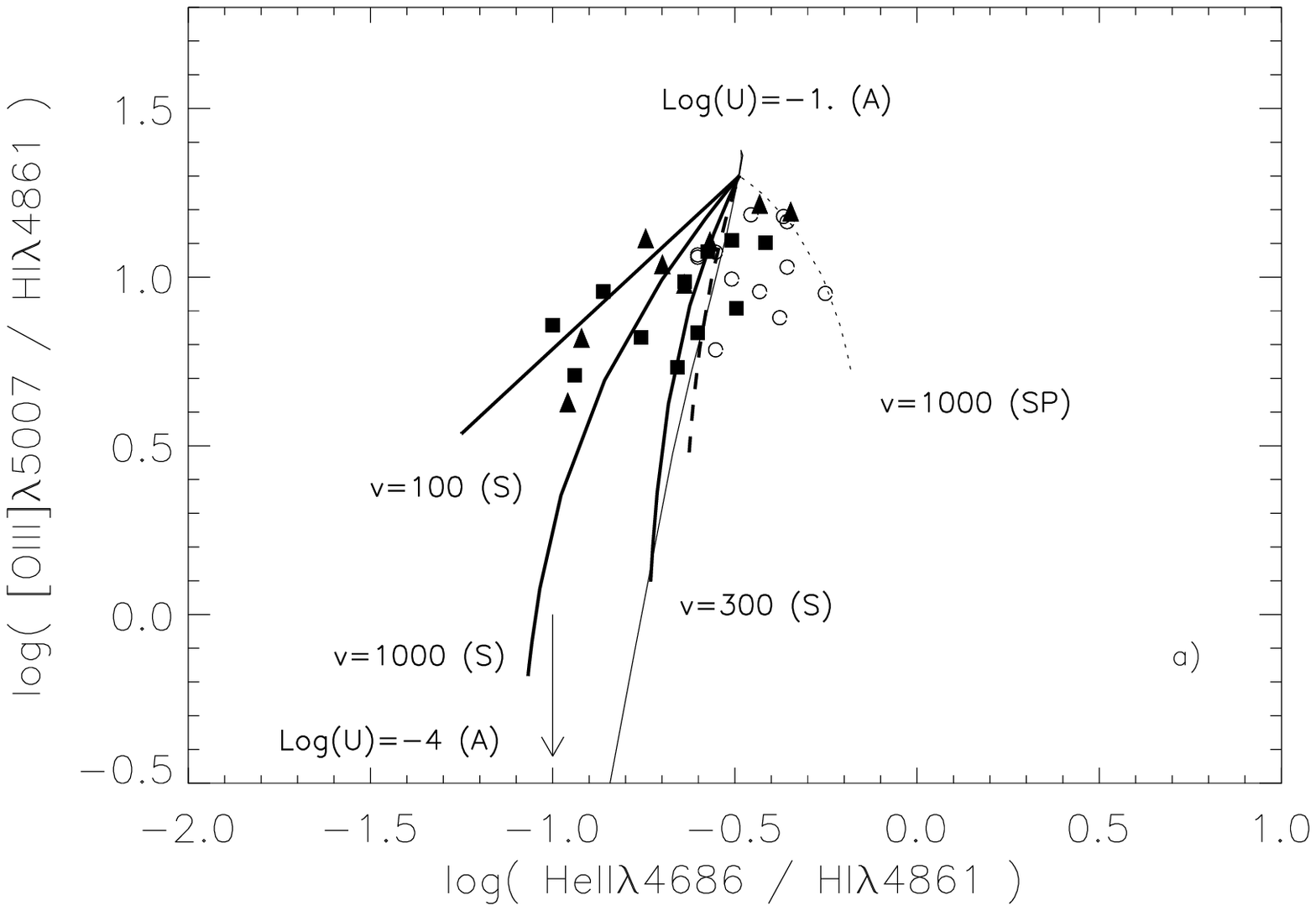,bbllx=30pt,bblly=0pt,bburx=480pt,bbury=
350pt,height=6cm}
\psfig{figure=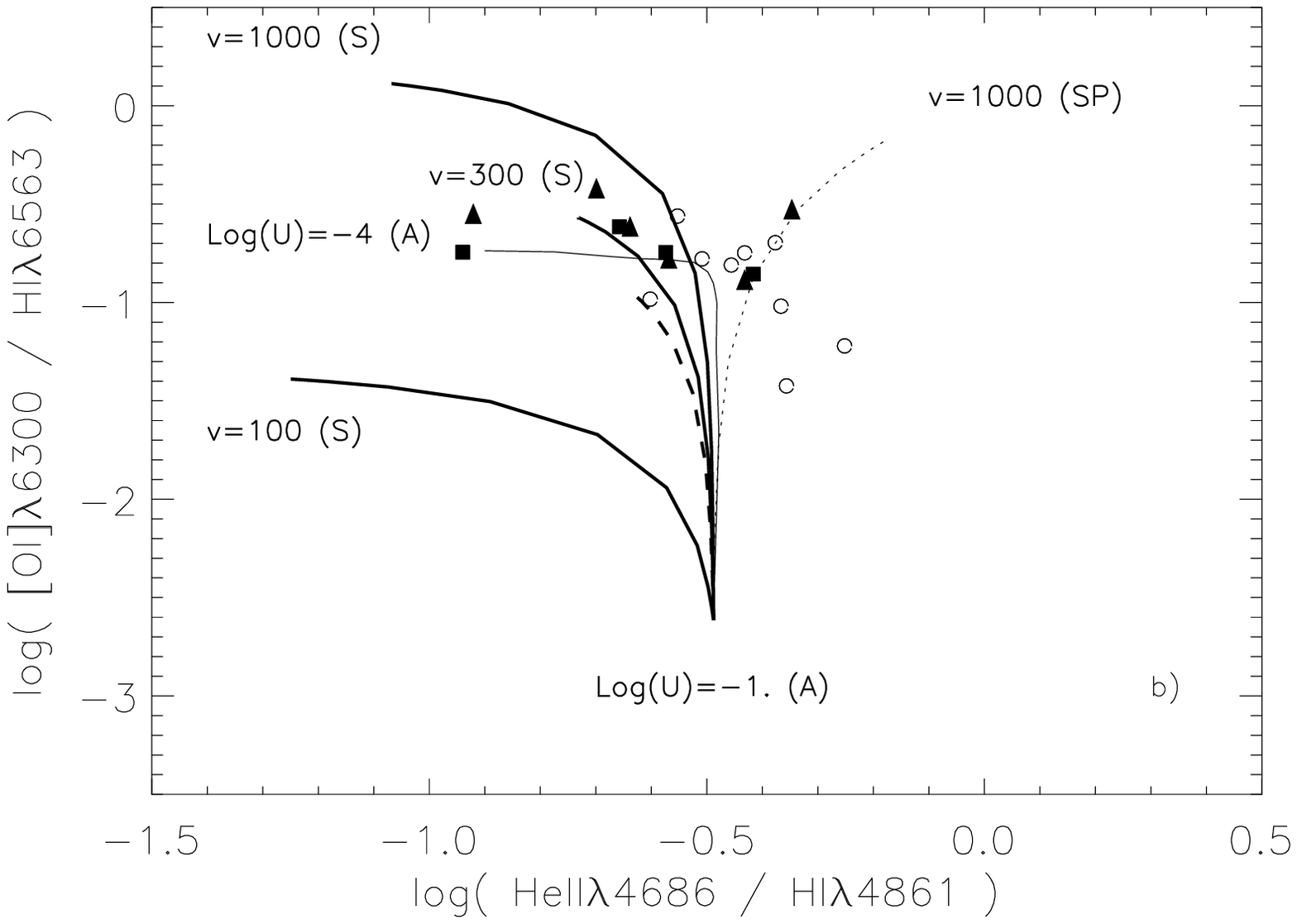,bbllx=0pt,bblly=0pt,bburx=480pt,bbury=
350pt,height=6cm}
}}
\parbox{14cm}{\caption{[OIII]$\lambda$5007/H$\beta$ and [OI]$\lambda$6300/H$\alpha$ vs. HeII$\lambda$4686/H$\beta$. Symbols  as in Fig. 1,  and curves as in
Fig. 2. (except for the dotted line; see below). Many extended regions (open circles) show high HeII$\lambda$4686/H$\beta$ ratios. They can be accounted for by A+SP models if the precursor is matter bounded (dotted line).}}
\end{center}
\end{figure*}
\end{center}

Most data points lie above log([OIII]$\lambda$5007/H$\beta$) = 0.5, and only five show a low ionization level (log([OIII]$\lambda$5007/H$\beta$) $\sim$ 0.). These objects (or regions) are probably completely dominated by shocks. The fact that there is no nuclear region  in this  group supports our initial hypothesis (Sect. 1) that shocks mainly occur far from the nucleus.

\subsubsection{The [OIII]$\lambda$4363/[OIII]$\lambda$5007 ratio}

The so-called ``temperature problem'' has long been an outstanding question in the modeling of emission lines in AGN. Photoionization models predict too low [OIII]$\lambda$4363/[OIII]$\lambda$5007 ratios (i.e., too low temperatures) compared to the observations. (e.g. Tadhunter et al. 1989). Dopita \& Sutherland (1995) showed that pure shock models are able to account for log([OIII]$\lambda$4363/[OIII]$\lambda$5007) as high as -1. However, these models underpredict the ionization level of the gas and are unable to account for the [OIII]$\lambda$5007/H$\beta$ ratios in AGN (see Fig. 7 in Dopita \& Sutherland 1995).

Figure 2 presents  the temperature-sensitive [OIII]$\lambda$4363/[OIII]$\lambda$5007 ratio vs. the two ionization level-sensitive  [OIII]$\lambda$5007/H$\beta$ and  [OI]$\lambda$6300/[OIII]$\lambda$5007 ratios (Fig. 2a and 2b, respectively). From these plots it is clear that A+S sequences solve the temperature problem. The bulk of the [OIII]$\lambda$4363 emission originates in shocks,  while the [OIII]$\lambda$5007 line is emitted mainly by the photoionized component.   An alternative to the A+S solution would be to increase the density, but we have shown in the previous section that a significant contamination of the optical lines by a high-density zone is unlikely.

The data distribution in the [OI]$\lambda$6300/[OIII]$\lambda$5007 vs. [OIII]$\lambda$4363/[OIII]$\lambda$5007 (Fig. 2b) is striking. In the classical AGN photoionization hypothesis, one would expect an inverse correlation between these two ratios, since [OIII]$\lambda$4363 weakens faster than [OIII]$\lambda$5007 at low $U$ (thin line). On the contrary, we find that [OI]$\lambda$6300/[OIII]$\lambda$5007  and [OIII]$\lambda$4363/[OIII]$\lambda$5007 are correlated. A Spearman's rank test gives a correlation coefficient of 0.41 with a significance level of $\sim$ 5 10$^{-7}$. This means that objects showing the {\it lowest} ionization level also have the {\it highest} electronic temperature. As illustrated in Fig. 2b, this trend is qualitatively reproduced by A+S models. Again, this result strongly suggests that the observational sequence from high-ionization to low-ionization radio sources is actually a sequence from photoionization-dominated to shock-dominated objects.

\begin{center}
\begin{figure*}[htbp]
\begin{center}
\centerline{\hbox{
\psfig{figure=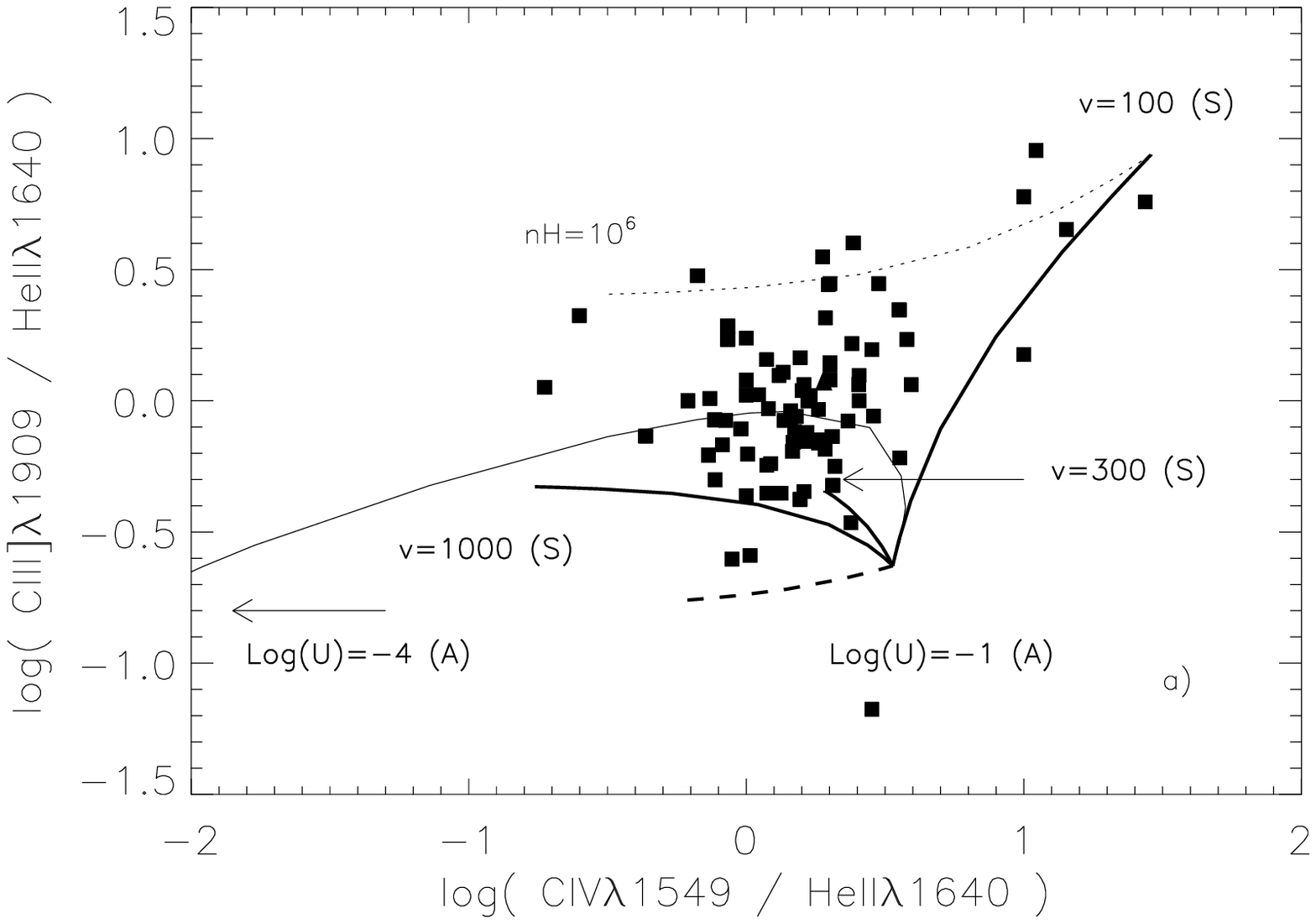,bbllx=30pt,bblly=0pt,bburx=480pt,bbury=
350pt,height=6.cm}
\psfig{figure=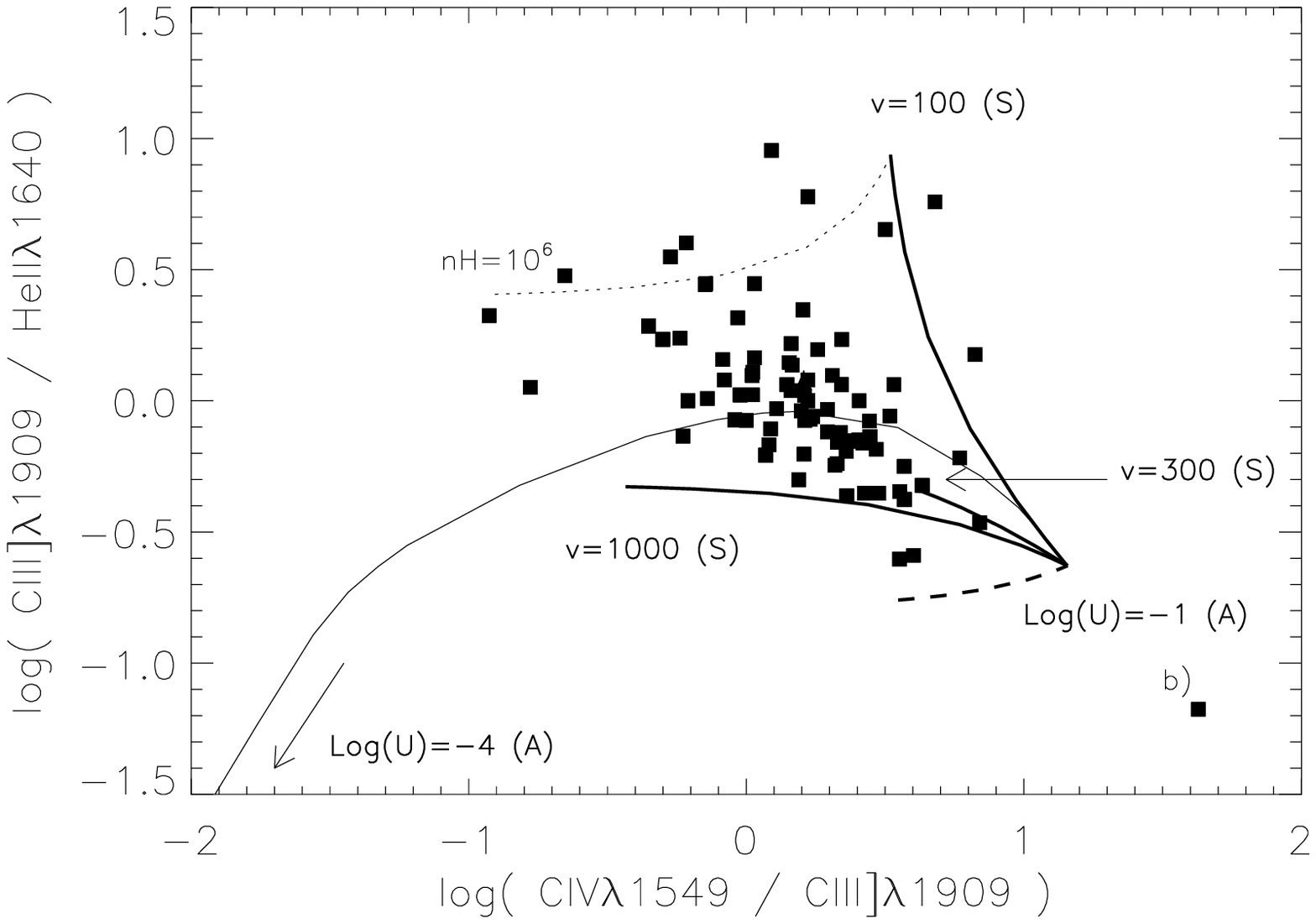,bbllx=0pt,bblly=0pt,bburx=480pt,bbury=
350pt,height=6.cm}
}}
\centerline{\hbox{
\psfig{figure=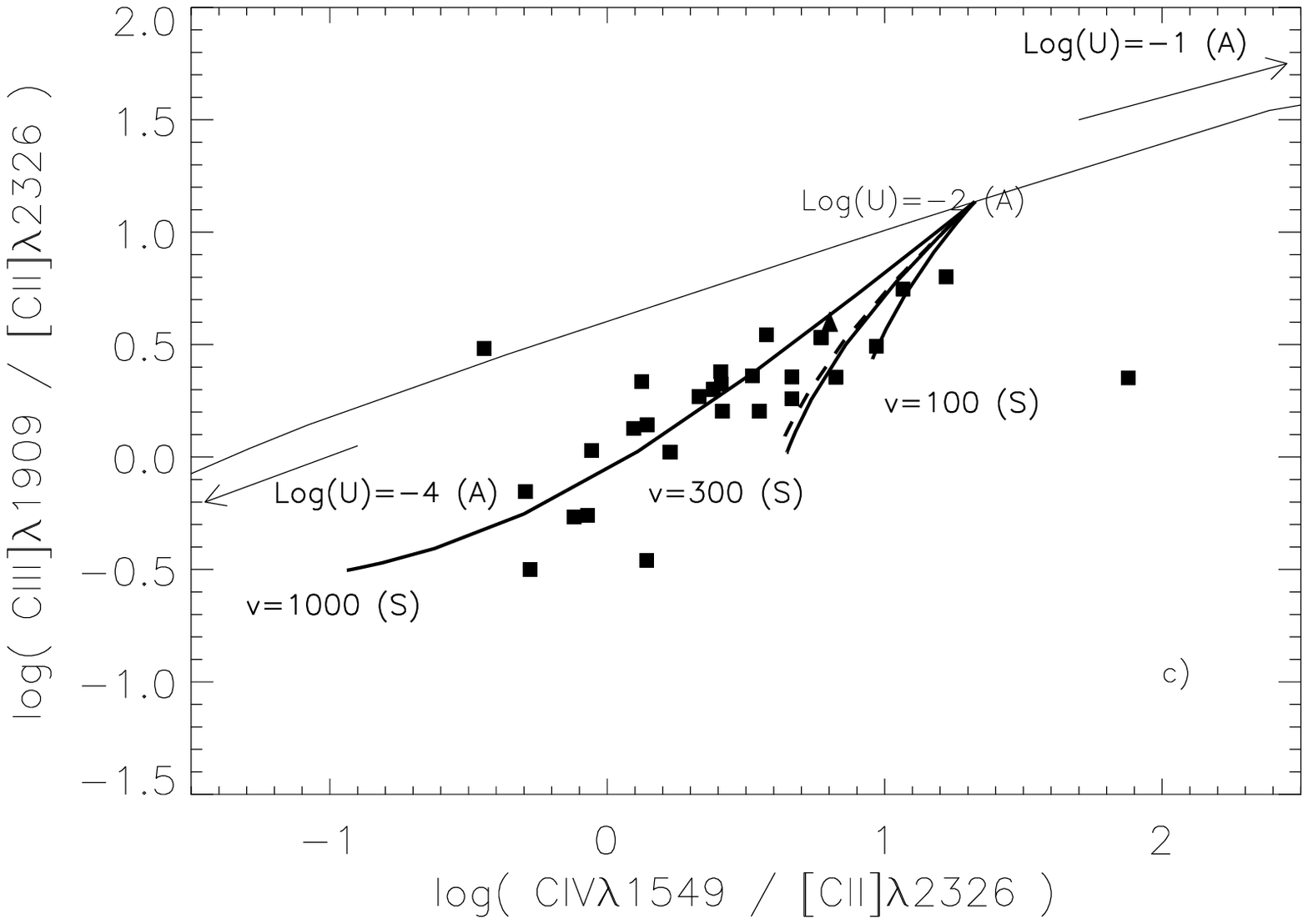,bbllx=30pt,bblly=0pt,bburx=480pt,bbury=
350pt,height=6.cm}
\psfig{figure=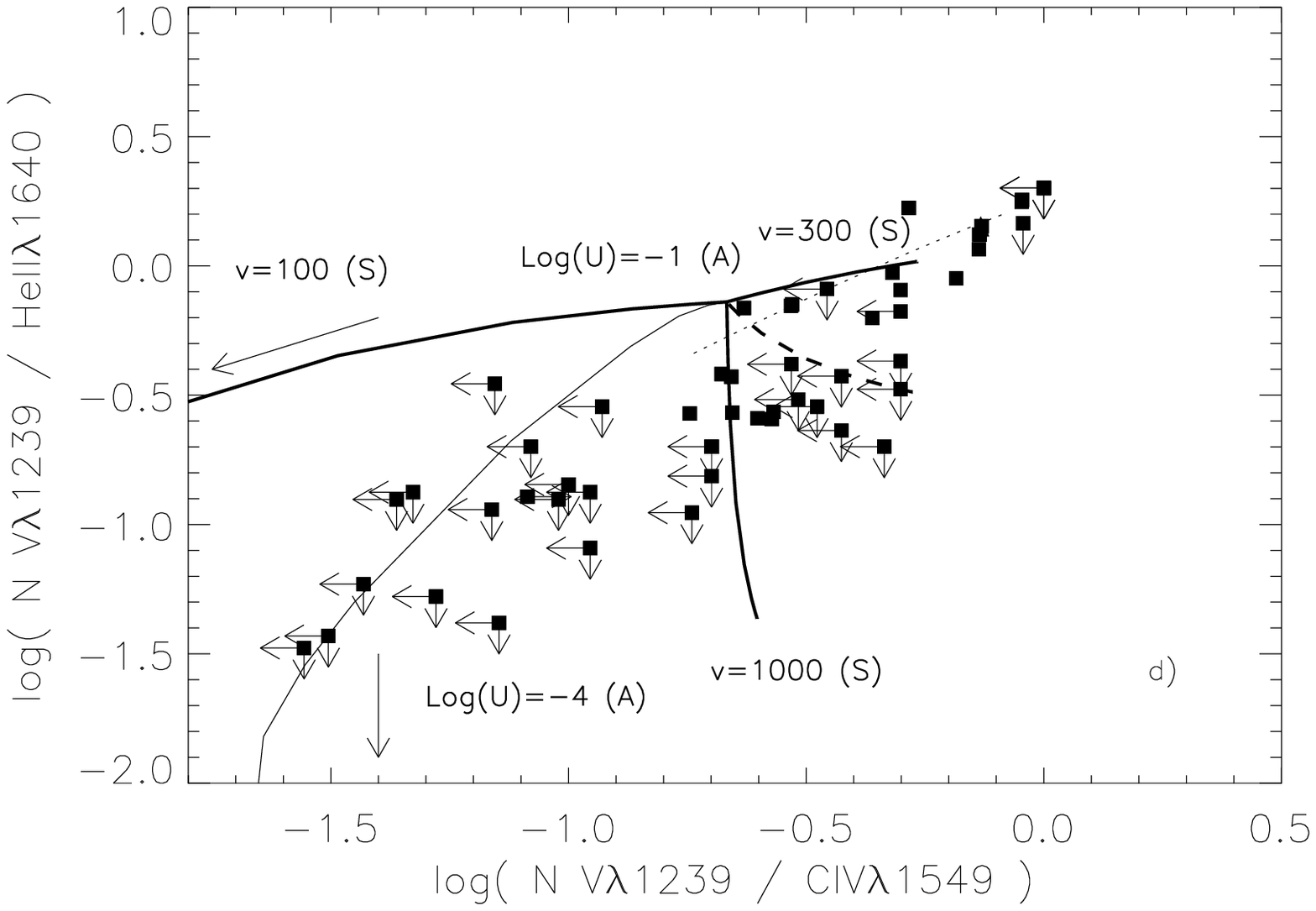,bbllx=0pt,bblly=0pt,bburx=480pt,bbury=
350pt,height=6.cm}
}}
\centerline{\hbox{
}}
\parbox{14cm}{\caption{Comparisons between the predictions of A+S and A+SP models and observational data in UV line ratios. Symbols as in Fig. 1. Thick solid curves, thin solid curves,  and dashed curve  are as in Fig. 2, except for Fig. 5c (see below). The dotted curve on Fig. 5a and 5b shows the A+S sequence with log\,$U$ = -3 $v$ = 100, if the density of the photoionized component is $n_\mathrm{H}$ = 10$^{6}$ cm$^{-3}$. The contamination of the emission line spectrum by the NLR is strongly suggested by these diagrams. In Fig. 5c, log\,$U$ = -1 has been replaced by log\,$U$ = -2 in the A+S sequence. In Fig. 5d, the dotted line shows the A+S sequence with log\,$U$ = -1 and $v$ = 300 km s$^{-1}$ for $Z$ = 1.5 $Z_{\odot}$.  }}
\end{center}
\end{figure*}
\end{center}

\subsubsection{[OI]$\lambda$6300/[OIII]$\lambda$5007 vs. [OII]$\lambda$3727/[OIII]$\lambda$5007}

 This diagram involves two line ratios which are sensitive to the ionization level of the gas. They are frequently used in the literature as a diagnostic of the ionization source of the gas (e.g. Baldwin, Phillips and Terlevich 1981). Again,  Fig. 3 illustrates:
\begin{itemize}
\item that a variation of the balance between shocks and AGN photoionization  quantitatively mimics a variation of the ionization parameter $U$ in classical line ratio diagrams; and

\item that  A+S and A+SP models are better able than pure photoionization models to account for the data. 

\end{itemize}

It is clear from this plot that there is a partial degeneracy between the ionization parameter of the photoionized component and the value of $A_\mathrm{A/S}$, and that log\,$U$ could be less than -1 in some objects.   Even so,  shocks are still needed to account for the  [OI]$\lambda$6300//[OIII]$\lambda$5007.

 Interestingly,  shock models predict  a relatively small range of log([OII]$\lambda$3727/[OIII]$\lambda$5007) whichis systematically higher than the predictions of AGN photoionization models, in contrast to  other line ratios highly influenced by shocks like [OI]$\lambda$6300/H$\alpha$, [NII]$\lambda$6584/H$\alpha$ and [SII]$\lambda$6720/H$\alpha$ (see Fig. 1). The [OII]$\lambda$3727/[OIII]$\lambda$5007 ratio is therefore an inambiguous indicator of the presence of ionizing shocks inside a given object.

\subsubsection{The HeII$\lambda$4686/H$\beta$ ratio}

 It is well known that the HeII$\lambda$4686/H$\beta$ ratio cannot be accounted for by classical photoionization models. Figure 4 shows that A+S and A+SP models (thick lines) account for most of the data data in the [OIII]$\lambda$5007/H$\beta$ vs. HeII$\lambda$4686/H$\beta$ (Fig. 4a) and [OI]$\lambda$6300/H$\alpha$ vs. HeII$\lambda$4686/H$\beta$ (Fig. 4b). However, these models still underpredict  HeII$\lambda$4686/H$\beta$ by  $\sim$ 0.5 dex for a few data (open circles), especially on Fig. 4a.  

Binette et al. (1996) showed that photoionized matter-bounded (MB) clouds  were one solution which could account for the highest HeII$\lambda$4686/H$\beta$ ratios observed in AGN. In this picture, however, MB clouds are necessarily located near the nucleus. This scenario is unlikely to work for our data since most of the problematic points  correspond to ``extended'' regions. As an alternative, we argue that MB {\it precursors} may well explain the existence of off-nuclear regions with high HeII$\lambda$4686/H$\beta$ ratios. Precursors are likely to lie far from the nucleus if, as our hypothesis predicts, shocks occur mainly near the radio hot spots. Moreover, their ionization level is high, thanks to the proximity of their ionization source (Dopita \& Sutherland 1996). Fig. 4 plots the prediction of the $v$ = 1000 km s$^{-1}$ SP model with a truncated precursor (see Sect. 2.2). We conclude that MB precursors, the existence of which is discussed in Sect. 5.1, are able to account for log(HeII$\lambda$4686/H$\beta$) $\sim$ -0.3 in extended regions.

\subsection{UV line ratios}

We have shown in the previous section that the coexistence of shocks and photoionization in AGN is strongly favored by analysis of optical line ratios. The distribution of the data in many diagrams is incompatible with pure photoionization models, while the A+S  sequence with log\,$U$= $\sim$ -1 and $v$ = 300 km s$^{-1}$ provides a good fit to the observations. The scattering
of the data around this sequence  is well accounted for by A+S sequences with   100 km s$^{-1}$ $\le$ v $\le$  1000 km s$^{-1}$. 
This result needs to be confirmed in the UV. However, such a study  is
 difficult due to the  possible contamination of   the  UV lines by the narrow line region (NLR), wich has densities of up to $n_\mathrm{H}$ = 10$^{6}$ cm$^{-3}$ (Villar-Mart\'{\i}n et al. 1999b). In what follows we show that the UV line fluxes observed in radio sources are compatible with the coexistence of AGN photoionization and shocks, if the contamination by the NLR is taken into account.

  Figure 5 presents the predictions  of the A+S and A+SP models compared to the UV data. An A+S sequence with a high density ($n_\mathrm{H}$ = 10$^{6}$  cm$^{-3}$)  photoionization component (log\,$U$ = -3) is also shown for comparison (dotted line) on the diagrams involving density-sensitive line ratios. The  data distributions are completely accounted for in the CIII]$\lambda$1909/HeII$\lambda$1640 vs. CIV$\lambda$1549/HeII$\lambda$1640 (Fig. 5a) and  CIII]$\lambda$1909/HeII$\lambda$1640 vs. CIV$\lambda$1549/CIII]$\lambda$1909 diagrams (Fig. 5b) only if contamination by a dense region is assumed.

We find, in good agreement with De Breuck et al. (2000), that the CIII]$\lambda$1909/[CII]$\lambda$2326 ratio is a good discriminant between shocks and AGN photoionization. This ratio is plotted in Fig. 5c as a function of CIV$\lambda$1549/[CII]$\lambda$2326.  Once more, two-component models appear to be the only way to account for the observations.  The $n_\mathrm{H}$ = 10$^{6}$ cm$^{-3}$ model is not plotted on Fig. 5c, since neither CIV$\lambda$1549/[CII]$\lambda$2326 nor CIII]$\lambda$1909/[CII]$\lambda$2326 is sensitive to density.

Figure 5d shows the results of two-component models for  NV$\lambda$1240/HeII$\lambda$1640 vs. NV$\lambda$1240/CIV$\lambda$1549. This diagram is often used as a metallicity diagnostic in AGN environments (Hamann \& Ferland 1993). The high  NV$\lambda$1240/HeII$\lambda$1640 and NV$\lambda$1240/CIV$\lambda$1549 ratios  in radio galaxies and quasars are frequently interpreted as evidence of largely super-solar metallicities,  not only in broad line regions (Hamann \& Ferland 1999), but also on  much larger scales (Villar-Mart\'{\i}n et al. 1999b, De Breuck et al. 2000). Fig. 5d illustrates that extreme metallicity effects are not necessarily needed to explain the intensity of the NV line.  Many  data points can be accounted for by  A+S and A+SP models at solar metallicity. The objects showing the most extreme NV$\lambda$1240/CIV$\lambda$1549 ratios can easily be accounted for by slightly increasing the metallicity ($Z$ = 1.5 $Z_{\odot}$) in A+S models (dotted line).

\section{Discussion}

\begin{center}
\begin{figure}[thbp]
\begin{center}
\centerline{\hbox{
\psfig{figure=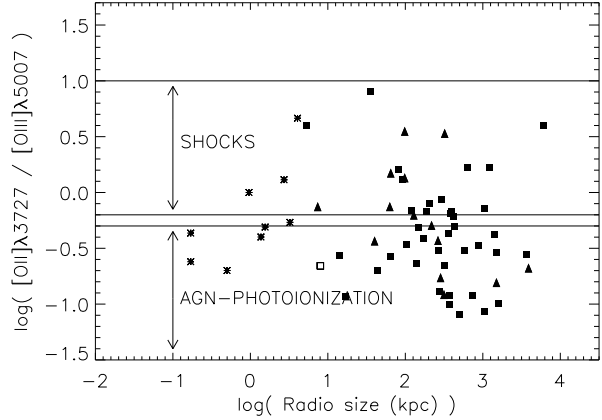,bbllx=0pt,bblly=0pt,bburx=550pt,bbury=
350pt,height=6cm}
}}
\parbox{8cm}{\caption{Evolution of the ionization-level sensitive [OII]/[OIII] ratio as a function of the radio size in radio galaxies. Filled squares represent FRII sources and  open squares FRI sources while stars represent  CSS sources. Triangles are the radio galaxies without known classification. The horizontal lines show the ranges of variation of the [OII]/[OIII] ratio in the cases of pure shocks and pure photoionization by the AGN. An interpretation of this diagram is given in Sect. 5.2 and 5.3.}}
\end{center}
\end{figure}
\end{center}

\subsection{Evidence for ionizing shocks in extended regions}

An extensive analysis of UV and optical line ratios in extragalactic radio sources has been presented in the previous sections. The results strongly support the hypothesis that the ionized gas in radio galaxies and QSOs is excited partly by the radiation from the central source and partly by shocks. First, the model sequences obtained by varying the shock contribution to the H$\beta$ flux ($A_\mathrm{A/S}$) between 0 and 100 \% provide  better fits to the data than classical $U$ sequences in all line ratio diagrams. Moreover, the observational sequence from high-ionization to low-ionization radio sources is naturally accounted for by varying  $A_\mathrm{A/S}$. The shock velocity lies between  100 km  s$^{-1}$ and 1000 km  s$^{-1}$. The ionization parameter log\,$U$ of the photoionized clouds is typically high, $\sim$  -1. \\
Our basic hypothesis (Sect. 1) was that ionizing shocks lie far from the nucleus inside radio sources. This assumption  is {\it a posteriori} justified by the high [OIII]$\lambda$5007/H${\beta}$ ratios observed in  {\it all} the central regions (Fig. 1), and by the fact that evidence for shocks is found mainly in extended regions. Another indirect indication of the prevalence of shocks far from the nucleus are the large HeII$\lambda$4686/H$\beta$ ratios observed in the external parts of radio galaxies (Fig. 4). These ratios can be accounted for only by models including a truncated (or ``matter-bounded'') precursor. The existence of high-$U$ MB clouds photoionized by the AGN (Binette et al. 1996) is unlikely so far from the central engine.  In contrast, the hypothesis  of MB precursors is reasonable. The ionizing photon source (the cooling post-shock medium) is very close. At large distances from the nucleus, it is possible that there is not enough material to absorb  all the ionizing photons emitted by the  post-shock zone. Finally, the hardness of the upward shock spectrum itself does not favor  100 \%  absorption of the ionizing photons.

\subsection{The radio size-ionizing mechanism relation}

 From Fig. 1 to 5, it is clear that the balance between shocks and AGN photoionization differs from one object to another. 
 Recently, Best et al. (2000b) presented convincing evidence that radio galaxies with radio sizes ($D$) larger than $\sim$ 150 kpc are dominated by AGN photoionization, and smaller sources by shocks. These results are based on the decrease of the [CII]$\lambda$2326/CIII]$\lambda$1909 ratio at large radio size, which has been  confirmed by De Breuck et al. (2000).  A confirmation  of the Best et al. (2000b) conclusions is clearly needed in the optical. The  ionization-level sensitive [OII]$\lambda$3727/[OIII]$\lambda$5007 ratio is ideal for such an analysis. The  [OII]$\lambda$3727/[OIII]$\lambda$5007 predictions are higher for shock models than for photoionization models (Fig. 3). In addition, [OII] and [OIII] fluxes   can easily be found from the literature, for objects covering a wide range of radio sizes,  permitting the construction of an ionization level-radio size  diagram covering four decades in $D$.   We plot on Fig. 6 the comparison between  [OII]/[OIII] and radio size for our sample. For comparison with Best et al. (2000b) and De Breuck et al. (2000),  as well as for the sake of homogeneity,  only the integrated data for radio galaxies, both FRI/II and CSS, are plotted.   Note that we have [OII] and [OIII] fluxes for 9 CSSs  (compared to the 23 CSSs included in our sample). The so-called ``weak line radio galaxies'' (e.g. Tadhunter et al. 1998) are not plotted,  due to uncertainty about the impact of the AGN inside these objects. Fig. 6 also shows the ranges of [OII]/[OIII] covered by photoionization models with $U$ $\ge$ 10$^{-3}$,  and by shock or shock+precursor models with 100 km s$^{-1}$ $\le$ $v$ $\le$ 1000 km s$^{-1}$.

Figure 6 confirms that there is a strong relation between the radio size and the ionization level of the gas. The most striking feature on Fig. 6 is the trend defined by CSS sources, namely a correlation between [OII]/[OIII] and $D$, although an {\it inverse} correlation is expected if shocks dominate in the most compact sources. Large reddening effects may explain this relation if the degree of reddening is a function of  radio size. Such a link is very unlikely, as demonstrated by Best et al. (2000b).  If CSS sources are frustrated by  a very dense medium (van Breugel et al. 1984), collisional de-excitation effects could explain the low [OII]/[OIII] in the CSS galaxies (since the critical density is lower for [OII] than for [OIII]. In principle, this could also explain the correlation between $D$ and [OII]/[OIII] if the most compact sources are confined by the densest media. 

To date, however, there is more support for the hypothesis that CSS are young rather than  frustrated (de Vries et al. 1998; Owsianik \& Conway 1998; Fanti et al. 2000;  Snellen et al. 2000). If it is the case, Fig. 6  strongly suggests that small radio sources are dominated by photoionization, then by shocks when they have grown up to $\sim$ 1 kpc, and finally by photoionization when $D$ $\ge$ $\sim$ 150 kpc. One exception to this scheme is 3C236. This galaxy, the largest FRII known ($D$ $\sim$ 6 Mpc), has a high [OII]/[OIII] ratio. This is possibly due to the existence inside 3C236 of an inner, compact (2 kpc) radio source (O'Dea et al. 2001). This source, presumably shock-dominated, may dominate the integrated emission of 3C236. \\

\subsection{Why are large and compact sources photoionization-dominated ?}

 We now investigate the origin of the relation between  radio size and shock-photoionization balance.  
 From Fig. 6,  compact/young sources seem to be dominated by photoionization.  The shock contribution then increases with radio size. This may be related to the increase of the cocoon surface (and hence of the shock area) as the hot spot advances, while the photoionization component remains constant. Best et al. (2000b) interpret the prevalence of photoionization  at larger radio sizes as a consequence of the  shock front having passed beyond the material surrounding the AGN.  Since most sources with $D$ $\ge$ 200 kpc are photoionization-dominated, this would imply that the warm gas around AGN typically  expands to a radius of $\sim$ 100 kpc from the nucleus.

These hypotheses about the increasing importance of ionizing shocks with time in small-scale radio sources, and their decreasing importance at larger scales, are in very good agreement with the evolution scenario of radio emission in Gigaherz Peaked Spectrum and FRIIs recently proposed by Snellen et al. (2000).

 Let us suppose that the ISM density  follows $n$ = $n_{0}$($x_{h}$/$x_{0}$)$^{-\delta}$, with $x_{0}$ = 1 kpc and $n_{0}$ = 100 cm$^{-3}$. The values of $\delta$ are typically $\sim$ 1.5 (Blundell et al. 1999). The H$\beta$ luminosity emitted by the shocks at the radio cocoon interface (Bicknell et al. 1997) depends on  $v_{3}$ = $v$/1000 km s$^{-1}$, and on the shock area $A_\mathrm{sh}$ as:

\begin{equation}
L_{\mathrm{H}\beta}^\mathrm{sh} = 1.91 \times 10^{-3} \ v_{3}^{2.41} \ n \ A_\mathrm{sh}
\end{equation}

\noindent The shock velocity $v$ depends on the hot spot-nucleus separation $x_{h}$ as:

\begin{equation}
v = v_{0}\left(\frac{x_{h}}{x_{0}}\right)^{(\delta-2)/3}
\end{equation}
 
\noindent where $v_{0}$ is a function of the jet energy flux $F_{E}$ and of the density $n_{0}$ following $v_{0}$ $\propto$ ($F_{E}$/$n_{0}$)$^{1/3}$. After elimination of $A_\mathrm{sh}$, Bicknell et al. (1997) show that the shock luminosity $L_{\mathrm{H}\beta}^\mathrm{sh}$  is given by:

\begin{equation}
L_{\mathrm{H}\beta}^\mathrm{sh} = 6.7 \ \times 10^{41} \ \left(\frac{6}{8-\delta}\right)^{0.8} \ F_{E}^{0.8} \ n^{0.2} \ \left(\frac{x_{h}}{x_{0}}\right)^{- (\delta-2)/5}
\end{equation}

With $\delta$ $\sim$ 1.5, the luminosity $L_{\mathrm{H}\beta}^\mathrm{sh}$ increases only slowly as the hot spot advances. In this case the expansion of the radio cocoon alone is insufficient to account for the  [OII]/[OIII]-radio size relation observed in CSSs. To solve this problem,  the photoionization contribution may progressively decrease during the expansion, for example due to the sweeping-out of material by the radio jet.

\section{Summary and conclusions}
In this paper we have presented strong evidence that both shocks and AGN photoionization  contribute to the ionization of the extended gas in  radio sources.  In ``classical'' optical line ratio diagrams, a variation between 0 and 100 \% of the shock contribution to the line emission mimics a variation of the ionization parameter of the gas, but the fit is better than with pure photoionization models. For [OI]$\lambda$6300/[OIII]$\lambda$5007 vs. [OIII]$\lambda$4363/[OIII]$\lambda$5007,   [OIII]$\lambda$5007/H$\beta$ vs. [OIII]$\lambda$4363/[OIII]$\lambda$5007  and   CIII]$\lambda$1909/[CII]$\lambda$2326 vs.  CIV$\lambda$1549/[CII]$\lambda$2326, the data can  be reproduced {\it only} with a mixture of shocks and AGN photoionization. 
The [OII]/[OIII] vs. radio size diagram shows that the most compact sources ($D$ $\le$ 2 kpc) and the sources larger than $D$ $\sim$ 150 kpc are dominated by AGN photoionization. Shocks dominate the ionizing process at {\it intermediate radio sizes}. This extends to small sizes the results of Best et al. (2000b) and De Breuck et al. (2001) obtained from UV line ratios. 

These results are in good agreement with the hypothesis that the radio cocoon and hot spots lead ionizing shocks into the ambient ISM. At the beginning of the expansion phase, the shock working surface is small and AGN photoionization prevails. Then shocks become progressively important as the working surface increases and/or the amount of material swept out by the jet reduces the importance of the photoionized component. Eventually the shock front passed beyond the edge of the ISM surrounding the nucleus and AGN photoionization prevails in  later phases.

High spatial resolution data are required to identify the shock-dominated zones  inside radio sources and confirm our results. Integral field spectrographs are clearly the ideal instruments for this purpose, mainly when mounted on 10-m class telescopes like the VLT.    

\begin{acknowledgements}
We would like to thank M. Dopita and L. Kewley for their help with using the MAPPINGSIII code,  C. O'Dea for useful discussions about the relation between the radio size and the dominant ionizing mechanism in AGN, and the anonymous referee for useful suggestions and comments. EM acknowledges support of the EU TMR Network ``Probing the Origin of the Extragalactic Background Radiation''  

\end{acknowledgements}

\newpage

\newpage

\appendix

\section{Source list}

\renewcommand{\arraystretch}{1.5}
\begin{table*}[htbp]
\begin{tabular}{|l|l|l|l|}
\hline
{\bf Reference$^{1}$} & {\bf Objects types$^{2}$} & {\bf Ref. for radio sizes$^{3}$} & {\bf Spatial covering} \\
\hline
Fosbury et al. (1987) & CSS & Tadhunter et al. (1994b) & Integrated\\
\hline
Robinson et al. (1987)  & RG & none & Nucleus \& Extended \\
\hline
Saunders et al. (1989) & RG &  3CR atlas $^{b}$ & Integrated\\
\hline
McCarthy et al. (1990) & RG & $^{a}$ & Integrated\\
\hline
Gelderman \&   Whittle (1994) & RG, CSS, QSO & $^{a}$ & Integrated \\  
\hline
Tadhunter et al. (1994a)  & RG & Krichbaum et al. (1998) & Extended\\
\hline
Soifer et al. (1995)$^{c}$ & QSO & none & Integrated \\
\hline
Simpson et al. (1996)  & RG & none & Integrated \\
\hline
Dey et al. (1997) & RG & none  & Integrated \\
\hline
Morganti et al. (1997) & CSS & $^{a}$ & Integrated \\
\hline
R\"ottgering et al. (1997) & RG & R\"ottgering et al. (1994) & Integrated \\
\hline
Clark et al. (1998)$^{d}$  & RG & Baum et al. (1988) & Integrated \\
\hline
Koekemoer \& Bicknell (1998)$^{e}$ & RG & Subrahmanyan et al. (1996) & Integrated \\
\hline
Lacy et al. (1998a)$^{f}$ & RG & none & Integrated \\
\hline
Lacy et al. (1998b) & QSO & Lawrence et al. (1993) &  Integrated \\
\hline
Tadhunter et al. (1998)  & RG, CSS, QSO & Morganti et al. (1993)$^{g}$ & Integrated \\
\hline
Villar-Mart\'{\i}n et al. (1998) & RG & $^{a}$ & Integrated\\
\hline
Best et al. (1999) & RG & $^{a}$ &  Integrated \\
\hline
Lacy et al. (1999) & RG, QSO& $^{a}$ & Integrated \\
\hline
Murayama et al. (1999) & QSO & none & Integrated \\
\hline
Simpson et al. (1999) & RG &  $^{a}$ & Integrated \\
\hline
Snellen et al. (1999) & RG, QSO& none & Integrated \\
\hline
Villar-Mart\'{\i}n et al (1999b) & RG & Pentericci et al (1999), & Integrated\\
                            &    & R\"ottgering et al. (1994) & \\
\hline
Villar-Mart\'{\i}n et al. (1999c) & RG & Morganti et al. (1993) & Extended\\
\hline
Best et al. (2000a)  & RG & $^{a}$ & Integrated \\
\hline
De Breuck et al. (2000) & RG &  $^{a}$ & Integrated\\ 
\hline
Palma et al. (2000)    & RG &  $^{a}$ & Integrated \\
\hline
Schoenmakers (2000)  & RG & $^{a}$ & Integrated\\
\hline
De Breuck et al. (2001) & RG &  $^{a}$ & Integrated\\ 
\hline
\end{tabular}
\parbox{15cm}{\caption{Col. 1: Reference for line fluxes. Col. 2: type(s) of object included in this source (RG: radio galaxies; CSS: compact steep spectrum sources; Sey: Seyfert galaxies). Col. 3: Reference(s) for radio size. Col. 4: spatial covering. (a): Reference for emission lines and radio sizes are identical. (b): 3CR atlas on line: http://www.jb.man.ac.uk/atlas/basic.html. (c): Source of the [OI]$\lambda$6300 flux in F10214+4724 (Lacy et al., 1998b). (d): Sum of the Nucleus and  EELR components. (e): Sum of the Center and Tail components. (f): Sum of the ``a'' and ``c'' components. (g) Radio sizes estimated from the radio maps. }}
\end{table*}
\end{document}